\documentclass[aps,prd,eqsecnum,amsmath,nofootinbib,preprintnumbers]{revtex4}%
\usepackage{graphicx}
\usepackage{xcolor}
\usepackage{float}
\definecolor{wred}{rgb}{0,0.618,0.0}
\definecolor{wblue}{rgb}{.0,0.0,0.618}
\definecolor{wgreen}{rgb}{.0,0.618,0.0}
\usepackage{tikz}
\usepackage{multirow}
\usetikzlibrary{decorations.pathmorphing,calc}
\usepackage{bm}
\usepackage[
colorlinks=true,
filecolor=black,
linkcolor=wblue,
citecolor=wgreen,
pdfstartview=FitH,
pdfpagemode=None,
bookmarksopen=true,
urlcolor=blue%{blue!80!black!80}
]{hyperref}
\begin{document}
	\preprint{\hfill {\small {1}}}
\title{New Topological Gauss-Bonnet Black Holes in Five Dimensions}
\author{Yuxuan Peng}
\email{yxpeng@alumni.itp.ac.cn}
\affiliation{Department of Physics, School of Science, East China University of Technology, Nanchang, Jiangxi 330013, P.R. China}
\affiliation{CAS Key Laboratory of Theoretical Physics, Institute of Theoretical Physics, Chinese Academy of Sciences, Beijing 100190, P.R. China}

\begin{abstract}
	We investigate vacuum static black hole solutions of Einstein-Gauss-Bonnet gravity with a negative cosmological constant in five dimensions.
	These are solutions with horizons of nontrivial topologies. 
	The first one possesses a horizon with the topology $S^1 \times H^2$, and a varying Gauss-Bonnet coupling constant $\alpha$. By looking into its thermodynamic properties, we find that its specific heat capacity with fixed volume is negative, therefore it is thermodynamically unstable.
	The second one is equipped with a so-called ``Sol-manifold'' as its horizon, and interestingly, the product of the Gauss-Bonnet coupling constant $\alpha$ and the cosmological constant $\Lambda$ is fixed.
	For the second solution, the total energy and entropy vanish.
	These results enlarge our knowledge of both topological black holes in higher dimensions and the property of higher curvature corrections of gravitational theories.
\end{abstract}

\maketitle

\section{introduction}
Black holes are simple and meanwhile complicated objects in nature.
The smooth event horizon of a stationary black hole can only have spherical topology, therefore they are simple.
However, this simplicity is ensured only when the spacetime dimension is $4$ and the dominant energy condition (DEC) is satisfied --- this is the topology theorem by Hawking in 1972 \cite{Hawking:1971vc}.
In the year 1994 Chru\'{s}ciel and Wald \cite{Chrusciel:1994tr} proposed a topology theorem without the condition of smoothness.
Black hole horizons can be complicated while those two conditions are violated, i.e. one can consider higher dimensions or break DEC.

In $5$ dimensions, there are not only the famous ``Myers-Perry'' black holes \cite{Myers:1986un} with spherical topology $S^3$, also ``black rings'' found by Emparan and Reall \cite{Emparan:2001wn} with topology $S^1\times S^2$.
Actually, these two topologies are the only possibilities for $5$ dimensional asymptotically flat stationary blackhole horizons according to the generalization \cite{Galloway:2005mf} to higher dimensions of Hawking's topology theorem.
In dimensions $D > 5$ there are much more complicated black hole horizons, including $S^{D-2}$ and $S^{1}\times S^{D-3}$ topologies \cite{Emparan:2008eg,Obers:2008pj}.

DEC can be broken by introducing the negative cosmological constant $\Lambda$.
The asymptotically anti-de Sitter (AdS, the maximally symmetric space with negative curvature) black holes are solutions with negative $\Lambda$ and they can have horionzs of $3$ types: sphere with positive curvature, torus with flat geometry and hyperbolic space with negative curvature \cite{Lemos:1994xp,Lemos:1995cm,Birmingham:1998nr}.
These $3$ types appear in arbitrary dimensions.
If one considers generic black holes that are not asymptotically AdS, with negative $\Lambda$, there are even black hole horizons with arbitrary genus $g>1$ \cite{Aminneborg:1996iz}. These are spacetimes locally equal to pure AdS spacetimes.
Moreover, some static, plane symmetric solutions and cylindrically symmetric solutions of Einstein-Maxwell equations with a negative cosmological constant are investigated in the paper \cite{Cai:1996eg} by Rong-Gen Cai and Yuan-Zhong Zhang.

It is interesting to explore the possibility of certain nontrivial horizon types in the presence of a negative $\Lambda$.
If certain types were proven to be forbidden, this might help establishing new topology theorems; if certain types turned out to exist, these would be novel discoveries worth noticing.
There is an important example, Ref. \cite{Cadeau:2000tj}, which obtained nontrivial black holes in $5$ dimensions.
These black holes have $3$-dimensional horizons of different types.
These types all belong to the eight $3$-dimensional ``model-geometries'' classified by Thurston \cite{Th1}.
They are: Euclidean $E^3$, spherical$S^3$, hyperbolic $H^3$, $S^1\times S^2$, $S^1\times H^2$ and the so-called ``Sol'', ``Nil'' and ``$SL_2 R$'' geometries.
They all admit homogeneous metrics.
For more details of these eight manifolds one can refer to \cite{Cadeau:2000tj,Th1,Sc1}.
These manifolds should be compact if one uses them to construct black hole horizons, and the topologies and the compactification of these eight manifolds are also described in \cite{Cadeau:2000tj}.
Its main results are Einsteinian solutions of black holes with ``Sol'' and ``Nil'' horizons\footnote{According to \cite{Cadeau:2000tj}, there had already been black hole horizons for the first five geometries except $S^1\times S^2$, while ``$SL_2 R$'' horizons are still unknown.}.
For recent devolopments of this 

What we do is to explore the effect of higher curvature corrections on horizon topology.
To be exact, we consider $5$-dimensional Einstein-Gauss-Bonnet (EGB for short) gravity theory, a special case of the Lovelock gravity theory \cite{Lovelock:1971yv}, the general second-order covariant gravity theory in dimensions higher than four, with a negative $\Lambda$. 
Moreover, the Gauss-Bonnet term can be regarded as corrections from the heterotic string theory \cite{Gross:1986iv,Zumino:1985dp}. 
There had already been spherical, Euclidean and hyperbolic black hole horizons in this theory, shown by the famous paper \cite{Cai:2001dz} by Rong-Gen Cai.
What we found are a black hole with a $S^1 \times H^2$ horizon and one with a Sol-manifold horizon.
These are static vacuum black holes.
We then discuss the thermodynamic properties of these solutions, and interestingly, the Sol-manifold solution has zero entropy and mass.
This may be caused by the fact that in the Sol-manifold solution the Gauss-Bonnet coupling constant has a fixed value.
We make some discussions at the end of this paper.
We hope our results may help further understanding black holes in higher curvature gravity theories and help exploring possible topology theorems in generic cases.

\section{The topological black hole solutions}
The action of EGB gravity theory with a negative cosmological constant
\begin{eqnarray}
S &= &\frac{1}{16 \pi G}\int \mathrm{d}^{d+1} x \sqrt{-g}\left[ R - 2\Lambda + \alpha (R^2 -4 R_{\mu\nu}R^{\mu\nu} + R_{\mu\nu \rho \sigma}R^{\mu\nu \rho \sigma}) \right]
%\nonumber\\
%&&-\frac{1}{4}\int \mathrm{d}^{d+1} x \sqrt{-g}F_{\mu\nu}F^{\mu\nu}
\,,
\end{eqnarray}
where $G$ is the Newton constant and the cosmological constant $\Lambda$, the Gauss-Bonnet coupling denoted by $\alpha$.
The equations of motion are
\begin{eqnarray}
&&R_{\mu\nu} - \frac{1}{2}R g_{\mu\nu} + \Lambda g_{\mu\nu}\nonumber\\
&&- \alpha\left(4 R_{\mu\rho} R^\rho\,_\nu - 2R R_{\mu\nu}+ 4 R^{\rho \sigma}R_{\mu\rho\nu\sigma} - 2 R_\mu\,^{\rho\sigma\tau}R_{\nu\rho\sigma\tau} + \frac{1}{2} g_{\mu\nu} (R^2 -4 R_{\rho \sigma}R^{\rho \sigma} + R_{\rho\sigma\tau\pi}R^{\rho\sigma\tau\pi})\right)\nonumber\\
&&=0
%&&\frac{1}{2}(F_{\mu\rho}F_\nu{}^{\rho}-\frac{1}{4}F_{\rho \sigma}F^{\rho \sigma}g_{\mu\nu})
\,.
\end{eqnarray}
%\begin{eqnarray}
%\nabla_\mu F^{\mu\nu}=0\,.\label{EMeq}
%\end{eqnarray}

\subsection{The first solution with a constant curvature subspace}

%A warped metric of $BH_3\times \Sigma_2$ ($BH_3$ is a $3$-dimensional static black hole with coordinates $t,r,\xi$ and $\Sigma_2$ is a $2$-dimensional constant curvature space) and a radial vector potential $A_{\mu} = A_t(r)\delta^t{}_\mu$
%give the equations of motion
%\begin{eqnarray}
%\bar{R}_{ab} - \frac{1}{2}(\bar{R}+\tilde{R})g_{ab} + \Lambda g_{ab} + \alpha \tilde{R}(2\bar{R}_{ab} - \bar{R} g_{ab})=\frac{1}{2}(F_{ac}F_b{}^{c}-\frac{1}{4}F_{cd}F^{cd}g_{ab})
%\end{eqnarray}
%and
%\begin{eqnarray}
%\tilde{R}_{ij} - \frac{1}{2}(\tilde{R}+\bar{R})g_{ij} + \Lambda g_{ij} + \alpha \bar{R}(2\tilde{R}_{ij} - \tilde{R} g_{ij})=-\frac{1}{8}F_{cd}F^{cd}g_{ij}\,,
%\end{eqnarray}
%where the quantities with a bar are the intrinsic quantities on $BH_3$ equipped with indices $a, b, ...$, while the tilde means the intrinsic quantities on $\Sigma_2$ with indices $i, j, ...$.
As we will see, the horizon topology of the first black hole solution is $S^1 \times H^2$.
The metric of the first solution is assumed to be a warped product of a $3$-dimensional static black hole $BH_3$ and a $2$-dimensional hyperbolic space $\Sigma_2$ with constant negative curvature:
\begin{eqnarray}
\mathrm{d}s^2 = -V{(r)}\mathrm{d}t^2 + \frac{1}{V(r)}\mathrm{d}r^2 + r^2\mathrm{d}\xi^2 + a \mathrm{d}\Sigma_2^2
\end{eqnarray}
with $V(r)$ an unknown function of coordinate $r$, $a$ a positive constant, and $\mathrm{d}\Sigma_2^2$ the line element of the $2$-d hyperbolic space,
\begin{eqnarray}
	\mathrm{d}\Sigma_2^2 = \mathrm{d}\theta^2 + \sinh^2 \theta \mathrm{d}\phi^2\,.
\end{eqnarray}
%Then we immediately have $A_t \sim \ln r$ from Eq.(\ref{EMeq}).
%The equations of motion are as follows:
%\begin{eqnarray}
%&&\text{$t-t$ component}:\qquad-\frac{V(r) \left(2 (a \Lambda  r+r)+(a-4 \alpha ) V'(r)\right)}{2 a r} = 0\\
%&&\text{$r-r$ component}:\qquad\frac{2 (a \Lambda  r+r)+(a-4 \alpha ) V'(r)}{2 a r V(r)} = 0\\
%&&\text{$\xi-\xi$ component}:\quad \frac{r^2 \left(2 a \Lambda +(a-4 \alpha ) V''(r)+2\right)}{2 a} = 0\\
%&&\text{$\theta-\theta$ component}:\quad a \Lambda +\frac{1}{2} a V''(r)+\frac{a V'(r)}{r} = 0\\
%&&\text{$\phi- \phi$ component}:\quad \frac{b \sinh ^2(\theta) \left(r \left(2 \Lambda +V''(r)\right)+2 V'(r)\right)}{2 r} = 0\,.
%\end{eqnarray}
The new black hole solution in $5$ dimensions has
%\begin{eqnarray}
%\mathrm{d}s^2 = - (-\frac{\Lambda}{3}r^2 - M)\mathrm{d}t^2 + \frac{1}{-\frac{\Lambda}{3}r^2 - M}\mathrm{d}r^2 + r^2 \mathrm{d}\xi^2 + (\frac{3}{-2\Lambda} - 2\alpha)\mathrm{d}\Sigma_2^2
%\end{eqnarray}
%with
\begin{eqnarray}
V(r)=-\frac{\Lambda  r^2}{3}-M\,,\qquad a=\frac{3}{-2\Lambda} - 2\alpha\,,
\label{solution1}
\end{eqnarray}
where $\alpha$ is the Gauss-Bonnet parameter, $\Lambda$ is the negative cosmological constant.
The constant $M$ is related to the total energy of the black hole, and is related to the horizon radius $r_h$ by the relation
\begin{eqnarray}
M = - \frac{\Lambda r_h^2}{3}\,.
\label{horizonradius}
\end{eqnarray}
The solution above with $\alpha\rightarrow0$ gives the Eq.(II.18) in the paper \cite{Cadeau:2000tj}.

\subsection{The second solution with nontrivial horizon topology}
The horizon of the second black hole solution is the so-called Sol-manifold\cite{Cadeau:2000tj,Th1,Sc1}.
The Sol-manifold is described by
\begin{eqnarray}\label{3dSolv}
\mathrm{d}s^2 = e^{2z}\mathrm{d}x^2 + e^{-2z}\mathrm{d}y^2 + \mathrm{d}z^2\,,
\end{eqnarray}
The ansatz metric of the whole spacetime is 
\begin{eqnarray}\label{Solansatz}
\mathrm{d}s^2 = -V{(r)}\mathrm{d}t^2 + \frac{1}{V(r)}\mathrm{d}r^2 +f(r)e^{2z}\mathrm{d}x^2 + g(r)e^{-2z}\mathrm{d}y^2 + h(r)\mathrm{d}z^2
\end{eqnarray}
which gives the metric on a constant $\{t,r\}$ surface up to some constant.
Here $V(r), f(r), g(r)$ and $h(r)$ are functions of $r$ to be determined.

After putting the ansatz into the equations of motion, we obtain a solution as follows:
\begin{eqnarray}
V(r)=-\frac{\Lambda  r^2}{3}-M\,,\qquad g(r)=r^2\,,\qquad h(r)=\frac{r^2}{M}\,.
\label{solution2}
\end{eqnarray}
Here the coupling constant $\alpha$ is fixed to satisfy
\begin{eqnarray}\label{criticalalpha1}
\alpha = -\frac{3}{4\Lambda}
\end{eqnarray}
and $M$ is an integration constant and $f(r)$ an unfixed function\footnote{One can also set $f(r)=r^2$, and leave $g(r)$ undetermined. These are the same solutions with different signs of $z$} even when the equations of motion are satisfied.
The horizon radius $r_h$ also satisfies the equation (\ref{horizonradius}).

This solution is quite different from the Sol-manifold solution of Einstein gravity in \cite{Cadeau:2000tj}.
Interestingly, when $f(r)$ and $g(r)$ are both set to be $r^2$, the horizon manifold can be arbitrary as shown in the paper \cite{Dotti:2007az}.
%The detailed process of obtaining this solution is given in the appendix.

\section{Thermodynamics of the nontrivial solutions}
In this section we study the thermodynamics of the black hole solutions given above.
\subsection{The first solution}
For the first solution (\ref{solution1}), we can identify the cosmological constant with the thermodynamic pressure $P = -\Lambda/(8\pi G)$. This identification had been applied to explore the extended phase space of asymptotically AdS black holes\cite{Kubiznak:2012wp}.

The temperature can be obtained by the semi-classical method of removing the conical singularity of the near-horizon geometry
\begin{eqnarray}
T=\frac{{V'(r)}}{4 \pi }\Big|_{r\to r_h} =-\frac{\Lambda  r_h}{6 \pi } = \frac{4 G P r_h}{3}\,,
\end{eqnarray}
and the entropy can be obtained by applying the Wald entropy formula \cite{Wald:1993nt,Iyer:1994ys}
\begin{eqnarray}\label{entropy}
S = -2\pi \int_{\text{horizon}} \sqrt{\hat{g}} \,\mathrm{d}^{d-1}x \frac{\partial L}{\partial R_{abcd}}{\bm \epsilon}_{ab}{\bm \epsilon}_{cd} = -\frac{3 r_h \Omega  (1+ 4 \alpha  \Lambda )}{8 G \Lambda } = \frac{3 r_h \Omega  (1-32 \pi  \alpha  G P)}{64 \pi  G^2 P}\,.\label{Waldentropy}
\end{eqnarray}
where all the hatted quantities are intrinsic quantities on the horizon cross section on which the integral is defined and ${\bm \epsilon}_{ab}$ is the natural volume element on the tangent space orthogonal to the cross section. The constant $\Omega$ is the volume of the hyperbolic subspace.

If we consider the first law of thermodynamics $\mathrm{d}E = T\mathrm{d}S$ without including $\Lambda$ or $P$ as a variable, we will obtain the expression for the total energy
\begin{eqnarray}
E = \frac{r_h^2 \Omega  (1 + 4 \alpha  \Lambda)}{32 \pi  G}\,.
\end{eqnarray}
However, when $\Lambda$ or $P$ is a thermodynamic variable, this quantity should be interpreted as the total enthalphy
\begin{eqnarray}\label{enthalpy}
H = \frac{r_h^2 \Omega  (1 + 4 \alpha  \Lambda)}{32 \pi  G} = \frac{3 r_h \Omega  (1-32 \pi  \alpha  G P)}{64 \pi  G^2 P}\,.
\end{eqnarray}
After rewriting $H(r_h, P)$ as the function of $S$ and $P$, we arrive at
\begin{eqnarray}
H(S,P) = \frac{128 \pi  G^3 P^2 S^2}{9 \Omega -288 \pi  \alpha  G P \Omega }\,,
\end{eqnarray}
and the first law of thermodynamics becomes
\begin{eqnarray}
\mathrm{d} H = T \mathrm{d} S + V\mathrm{d} P
\end{eqnarray}
with $V$ being the thermodynamic volume
\begin{eqnarray}
V = \frac{256 \pi  G^3 P S^2 (1-16 \pi  \alpha  G P)}{9 \Omega  (1-32 \pi  \alpha  G P)^2} = \frac{1}{16} r_h^2 \Omega  \left(\frac{1}{\pi  G P}-16 \alpha \right) = \frac{9 T^2 \Omega  (1-16 \pi  \alpha  G P)}{256 \pi  G^3 P^3}\,.
\end{eqnarray}
The last equation above gives the equation of state of the black hole.
One can see that the derivative of $V$ with respect to $P$ is always negative
\begin{eqnarray}
\left(\frac{\partial V}{\partial P}\right)_T = \frac{9 T^2 \Omega  (32 \pi  \alpha  G  P-3)}{256 \pi  G^3 P^4} <0
\end{eqnarray}
for all $T$, since $1-32 \pi  \alpha  G P>0$ must be satisfied in Eq.(\ref{entropy}).
%Therefore, This black hole does not show $P-V$ criticality.

The specific heat capacity $C_p$ with fixed pressure is
\begin{eqnarray}
C_p = T \left(\frac{\partial S}{\partial T}\right)_P = \frac{9 T \Omega  (1-32 \pi  \alpha  G P)}{256 \pi  G^3 P^2} > 0
\end{eqnarray}
and the specific heat capacity $C_v$ with volume fixed is
\begin{eqnarray}
C_v=C_p - T\left(\frac{\partial P}{\partial T}\right)_V\left(\frac{\partial V}{\partial T}\right)_P = \frac{9 T \Omega }{256 \pi  G^3 P^2 (32 \pi  \alpha  G P-3)} < 0
\end{eqnarray}
since $1-32 \pi  \alpha  G P>0$.
So this system is thermodynamically unstable in the sense of specific heat capacity with fixed volume, $C_v$.
\subsection{The second solution}
After applying the Wald entropy (\ref{Waldentropy}) formula to the second solution (\ref{solution2}), we found that the entropy vanishes, since the integrand
\begin{eqnarray}
\frac{\partial L}{\partial R_{abcd}}{\bm \epsilon}_{ab}{\bm \epsilon}_{cd} \propto \frac{\left(3 M+\Lambda  r^2\right) f'(r)}{2 \Lambda  r f(r)}
\label{Solentropy}
\end{eqnarray}
vanishes on the horizon, i.e. at the point $r=r_h$.
According to the first law of thermodynamics $\mathrm{d}E = T\mathrm{d}S$, the total energy\footnote{For the second solution we do not address the extended phase space, so we only worry about the total energy instead of the enthalpy.} of this solution also vanishes as the entropy,
\begin{eqnarray}
S=0\,,\qquad E=0\,.
\end{eqnarray}
This implies that the integration constant $M$ in this solution does not stand for the total energy or mass.

\section{conclusion and discussion}
In this paper we present two novel $5$-dimensional black hole solutions in Einstein-Gauss-Bonnet theory with a negative cosmological constant.
They are simple vacuum static black holes with horizons with nontrivial horizons.
Their horizons corresponds to two of the $8$ types of $3$-dimensional ``Thurston model geometries'' in the literature\cite{Sc1,Th1}.

The horizon topology of the first black hole solution is $S^1 \times H^2$, and it allows an extended phase interpretation of thermodynamics by including the cosmological constant as the thermodynamic pressure. 
We derive the heat capacities of this solution and find that it is thermodynamically unstable.
This solution has the form of the so-called ``warped product'', the perturbations of which can be studied in the formalism provided by the paper \cite{Cai:2013cja} by Rong-Gen Cai and Li-Ming Cao.
In this formalism, the hyperbolicity and causality of this solution can be investigated in the same way as the paper \cite{Cao:2021sty} by Li-Ming Cao and Liang-Bi Wu. In that paper they analyzed various spacetimes, including dynamical spacetimes.

The second black hole has the ``Sol-manifold'' as its horizon.
The Gauss-Bonnet coupling constant $\alpha$ of this solution is fixed as $\alpha = -{3}/({4\Lambda})$.
Curiously, there is an unfixed function $f(r)$ in the metric in Eq.(\ref{Solansatz}).
It seems that there may be redundant degrees of freedom in this solution.
Moreover, by applying the Wald entropy formula we find that its entropy vanishes, so does its total energy due to the first law of thermodynamics.
This phenomenon is thought-provoking.
In the paper \cite{Cai:2009de} by Rong-Gen Cai, Li-Ming Cao and Nobuyoshi Ohta, some black hole solutions with zero mass and entropy had been presented in the context of Lovelock gravity theory.
The authors fix the coupling constant to such a critical value that the kinetic fluctuations around the background spacetime vanish, so there are no excitations of the background.
Since the entropy is related to the quantum degrees of freedom of the black hole, if the fluctuations are forbidden, the entropy is expected to vanish.
This is why the entropy vanishes in their case.
This is closely related to the fact that the effective gravitational constant $G_{\text{eff}}\rightarrow \infty$ \cite{Brustein:2007jj}.
Our case is similar: the coupling constant $\alpha$ is fixed, and the entropy is zero.
However, our coefficient choice is different from that paper. 
This is an intriguing fact, implying that there might be more than one critical value.
Our coefficient choice (\ref{criticalalpha1}) is the same as the paper on dimensionally continued gravity\cite{Banados:1993ur} and the paper \cite{Fan:2016zfs} by Zhong-Ying Fan, Bin Chen and Hong L\"u, where they found the critical value of the coupling constant actually forbids kinetic fluctuations around the AdS backgrounds.
The connection of the entropy and the kinetic fluctuations is quite interesting and deserves further investigation.

Among the $8$ model geometries\cite{Th1,Sc1} in $3$ dimensions, the black holes with $S^3\,, H^3$ and $R^3$ horizons already exist in both Einstein and Gauss-Bonnet gravity.
However, except the $S^1 \times H^2$ and the Sol-geometry horizons, we have not find static black holes with $S^1\times S^2$, Nil-geometry or $SL_2 R$ horizons in Gauss-Bonnet gravity yet, although such black holes (except $SL_2 R$) have been found in Einstein gravity.
To try to find or to rule out these solutions are important future directions.
To generalize our results to charged cases is also interesting.
%\footnote{An example of adding fields in the solution is the paper \cite{Naderi:2019jhn} in which the dolutions contain seven homogeneous Thurston’s geometries.}

Another interesting direction is about the connection between quantum information and gravity. The concept quantum complexity denotes the computation cost of reaching a certain quantum state, its holographic dual \cite{Stanford:2014jda,Brown:2015bva,Brown:2015lvg,Carmi:2016wjl} as well as its precise definition (still unclear) on the quantum theory side had received much attention in recent years\cite{Yang:2016awy,Carmi:2017jqz,Kim:2017qrq,Yang:2017nfn,Kim:2017lrw,Miao:2017quj,Yang:2018nda,Chen:2018mcc,An:2018dbz,Guo:2018kzl,Yang:2018tpo,Yang:2018cgx,Yang:2019gce,Bernamonti:2019zyy,Yang:2019iav,Yang:2019udi,Caceres:2019pgf,An:2019opz,Yang:2019alh,Bernamonti:2020bcf,Ruan:2020vze,An:2020tkn,Hernandez:2020nem,Yang:2020tna}.
The late-time growth rate of complexity $C$ is proportional to the product of the temperature and entropy $\dot{C} =T S$.
It will be interesting to look into the solutions with vanishing entropy and see the behavior of the complexity.

\section{Acknowledgements}
I should especially give my thanks to Prof. Rong-Gen Cai. Without him this work couldn't be done.
I also would like to thank Prof. Li-Ming Cao, Li Li, Yu-Sen An and Liang-Bi Wu for helpful discussions.
This work was supported in part by the National Natural Science Foundation of China with Grant No. 11947029, the National Postdoctoral Program for Innovative Talents with Grant No. BX201700259 and the ECUT launching program for Doctors with Grant No. DHBK2019198.


\begin{thebibliography}{99}
	


%\cite{Hawking:1971vc}
\bibitem{Hawking:1971vc}
S.~W.~Hawking,
%``Black holes in general relativity,''
Commun. Math. Phys. \textbf{25}, 152-166 (1972)
doi:10.1007/BF01877517
%913 citations counted in INSPIRE as of 13 Dec 2020
		

%\cite{Chrusciel:1994tr}
\bibitem{Chrusciel:1994tr}
P.~T.~Chrusciel and R.~M.~Wald,
%``On the topology of stationary black holes,''
Class. Quant. Grav. \textbf{11}, L147-L152 (1994)
doi:10.1088/0264-9381/11/12/001
[arXiv:gr-qc/9410004 [gr-qc]].
%89 citations counted in INSPIRE as of 13 Dec 2020		
		
		
%\cite{Myers:1986un}
\bibitem{Myers:1986un}
R.~C.~Myers and M.~J.~Perry,
%``Black Holes in Higher Dimensional Space-Times,''
Annals Phys. \textbf{172}, 304 (1986)
doi:10.1016/0003-4916(86)90186-7
%1787 citations counted in INSPIRE as of 13 Dec 2020		
		

%\cite{Emparan:2001wn}
\bibitem{Emparan:2001wn}
R.~Emparan and H.~S.~Reall,
%``A Rotating black ring solution in five-dimensions,''
Phys. Rev. Lett. \textbf{88}, 101101 (2002)
doi:10.1103/PhysRevLett.88.101101
[arXiv:hep-th/0110260 [hep-th]].
%869 citations counted in INSPIRE as of 13 Dec 2020

%\cite{Galloway:2005mf}
\bibitem{Galloway:2005mf}
G.~J.~Galloway and R.~Schoen,
%``A Generalization of Hawking's black hole topology theorem to higher dimensions,''
Commun. Math. Phys. \textbf{266}, 571-576 (2006)
doi:10.1007/s00220-006-0019-z
[arXiv:gr-qc/0509107 [gr-qc]].
%186 citations counted in INSPIRE as of 13 Dec 2020

%\cite{Emparan:2008eg}
\bibitem{Emparan:2008eg}
R.~Emparan and H.~S.~Reall,
%``Black Holes in Higher Dimensions,''
Living Rev. Rel. \textbf{11}, 6 (2008)
doi:10.12942/lrr-2008-6
[arXiv:0801.3471 [hep-th]].
%563 citations counted in INSPIRE as of 13 Dec 2020

%\cite{Obers:2008pj}
\bibitem{Obers:2008pj}
N.~A.~Obers,
%``Black Holes in Higher-Dimensional Gravity,''
Lect. Notes Phys. \textbf{769}, 211-258 (2009)
doi:10.1007/978-3-540-88460-6\_6
[arXiv:0802.0519 [hep-th]].
%47 citations counted in INSPIRE as of 13 Dec 2020

%\cite{Lemos:1994xp}
\bibitem{Lemos:1994xp}
J.~P.~S.~Lemos,
%``Cylindrical black hole in general relativity,''
Phys. Lett. B \textbf{353}, 46-51 (1995)
doi:10.1016/0370-2693(95)00533-Q
[arXiv:gr-qc/9404041 [gr-qc]].
%318 citations counted in INSPIRE as of 13 Dec 2020

%\cite{Lemos:1995cm}
\bibitem{Lemos:1995cm}
J.~P.~S.~Lemos and V.~T.~Zanchin,
%``Rotating charged black string and three-dimensional black holes,''
Phys. Rev. D \textbf{54}, 3840-3853 (1996)
doi:10.1103/PhysRevD.54.3840
[arXiv:hep-th/9511188 [hep-th]].
%228 citations counted in INSPIRE as of 13 Dec 2020

%\cite{Birmingham:1998nr}
\bibitem{Birmingham:1998nr}
D.~Birmingham,
%``Topological black holes in Anti-de Sitter space,''
Class. Quant. Grav. \textbf{16}, 1197-1205 (1999)
doi:10.1088/0264-9381/16/4/009
[arXiv:hep-th/9808032 [hep-th]].
%399 citations counted in INSPIRE as of 13 Dec 2020

%\cite{Aminneborg:1996iz}
\bibitem{Aminneborg:1996iz}
S.~Aminneborg, I.~Bengtsson, S.~Holst and P.~Peldan,
%``Making anti-de Sitter black holes,''
Class. Quant. Grav. \textbf{13}, 2707-2714 (1996)
doi:10.1088/0264-9381/13/10/010
[arXiv:gr-qc/9604005 [gr-qc]].
%198 citations counted in INSPIRE as of 13 Dec 2020


%\cite{Cai:1996eg}
\bibitem{Cai:1996eg}
R.~G.~Cai and Y.~Z.~Zhang,
%``Black plane solutions in four-dimensional space-times,''
Phys. Rev. D \textbf{54}, 4891-4898 (1996)
doi:10.1103/PhysRevD.54.4891
[arXiv:gr-qc/9609065 [gr-qc]].
%265 citations counted in INSPIRE as of 02 Jun 2021









	
		
%\cite{Cadeau:2000tj}
\bibitem{Cadeau:2000tj}
C.~Cadeau and E.~Woolgar,
%``New five-dimensional black holes classified by horizon geometry, and a Bianchi VI brane world,''
Class. Quant. Grav. \textbf{18}, 527-542 (2001).
%34 citations counted in INSPIRE as of 04 Oct 2020		

\bibitem{Th1} W.P.\ Thurston, {\it Three-Dimensional Geometry and 
	Topology}, ed.\ S.\ Levy (Princeton University Press, Princeton, 1997).

\bibitem{Sc1} P.\ Scott, {\it Bull.\ London Math.\ Soc.}\ {\bf 15} (1983),
401. 








		%\cite{Lovelock:1971yv}
\bibitem{Lovelock:1971yv} 
D.~Lovelock,
%``The Einstein tensor and its generalizations,''
J.\ Math.\ Phys.\  {\bf 12}, 498 (1971).
doi:10.1063/1.1665613
%%CITATION = doi:10.1063/1.1665613;%%
%1364 citations counted in INSPIRE as of 28 Apr 2018


%\cite{Gross:1986iv}
\bibitem{Gross:1986iv} 
D.~J.~Gross and E.~Witten,
%``Superstring Modifications of Einstein's Equations,''
Nucl.\ Phys.\ B {\bf 277}, 1 (1986).
doi:10.1016/0550-3213(86)90429-3
%%CITATION = doi:10.1016/0550-3213(86)90429-3;%%
%729 citations counted in INSPIRE as of 28 Apr 2018

%\cite{Zumino:1985dp}
\bibitem{Zumino:1985dp} 
B.~Zumino,
%``Gravity Theories in More Than Four-Dimensions,''
Phys.\ Rept.\  {\bf 137}, 109 (1986).
doi:10.1016/0370-1573(86)90076-1
%%CITATION = doi:10.1016/0370-1573(86)90076-1;%%
%392 citations counted in INSPIRE as of 28 Apr 2018	

		%\cite{Cai:2001dz}
\bibitem{Cai:2001dz} 
R.~G.~Cai,
%``Gauss-Bonnet black holes in AdS spaces,''
Phys.\ Rev.\ D {\bf 65}, 084014 (2002)
doi:10.1103/PhysRevD.65.084014
[hep-th/0109133].
%%CITATION = doi:10.1103/PhysRevD.65.084014;%%
%581 citations counted in INSPIRE as of 06 Mar 2018


%\cite{Dotti:2007az}
\bibitem{Dotti:2007az}
G.~Dotti, J.~Oliva and R.~Troncoso,
%``Exact solutions for the Einstein-Gauss-Bonnet theory in five dimensions: Black holes, wormholes and spacetime horns,''
Phys. Rev. D \textbf{76}, 064038 (2007)
doi:10.1103/PhysRevD.76.064038
[arXiv:0706.1830 [hep-th]].
%102 citations counted in INSPIRE as of 03 Jun 2021



		
%\cite{Kubiznak:2012wp}
\bibitem{Kubiznak:2012wp}
D.~Kubiznak and R.~B.~Mann,
%``P-V criticality of charged AdS black holes,''
JHEP \textbf{07}, 033 (2012).
%625 citations counted in INSPIRE as of 04 Oct 2020	


%\cite{Wald:1993nt}
\bibitem{Wald:1993nt} 
R.~M.~Wald,
%``Black hole entropy is the Noether charge,''
Phys.\ Rev.\ D {\bf 48}, no. 8, R3427 (1993).
%%CITATION = doi:10.1103/PhysRevD.48.R3427;%%
%1589 citations counted in INSPIRE as of 14 Mar 2020		
		
%\cite{Iyer:1994ys}
\bibitem{Iyer:1994ys} 
V.~Iyer and R.~M.~Wald,
%``Some properties of Noether charge and a proposal for dynamical black hole entropy,''
Phys.\ Rev.\ D {\bf 50}, 846 (1994).
%%CITATION = doi:10.1103/PhysRevD.50.846;%%
%1358 citations counted in INSPIRE as of 14 Mar 2020
		
		
		
		
		
%%\cite{Ishibashi:2011ws}
%\bibitem{Ishibashi:2011ws}
%A.~Ishibashi and H.~Kodama,
%%``Perturbations and Stability of Static Black Holes in Higher Dimensions,''
%Prog. Theor. Phys. Suppl. \textbf{189}, 165-209 (2011)
%doi:10.1143/PTPS.189.165
%[arXiv:1103.6148 [hep-th]].
%%64 citations counted in INSPIRE as of 04 May 2021


%\cite{Cai:2013cja}
\bibitem{Cai:2013cja}
R.~G.~Cai and L.~M.~Cao,
%``Generalized Formalism in Gauge-Invariant Gravitational Perturbations,''
Phys. Rev. D \textbf{88}, 084047 (2013)
doi:10.1103/PhysRevD.88.084047
[arXiv:1306.4927 [gr-qc]].
%7 citations counted in INSPIRE as of 03 May 2021		


%%\cite{Reall:2014pwa}
%\bibitem{Reall:2014pwa}
%H.~Reall, N.~Tanahashi and B.~Way,
%%``Causality and Hyperbolicity of Lovelock Theories,''
%Class. Quant. Grav. \textbf{31}, 205005 (2014)
%doi:10.1088/0264-9381/31/20/205005
%[arXiv:1406.3379 [hep-th]].
%%64 citations counted in INSPIRE as of 04 May 2021

%\cite{Cao:2021sty}
\bibitem{Cao:2021sty}
L.~M.~Cao and L.~B.~Wu,
%``Hyperbolicity and Causality of Einstein-Gauss-Bonnet Gravity in Warped Product Spacetimes,''
Phys. Rev. D \textbf{103}, no.6, 064054 (2021)
doi:10.1103/PhysRevD.103.064054
[arXiv:2101.02461 [gr-qc]].
%1 citations counted in INSPIRE as of 03 May 2021		
		

		
%\cite{Cai:2009de}
\bibitem{Cai:2009de}
R.~G.~Cai, L.~M.~Cao and N.~Ohta,
%``Black Holes without Mass and Entropy in Lovelock Gravity,''
Phys. Rev. D \textbf{81}, 024018 (2010)
doi:10.1103/PhysRevD.81.024018
[arXiv:0911.0245 [hep-th]].
%34 citations counted in INSPIRE as of 29 Nov 2020		


%\cite{Brustein:2007jj}
\bibitem{Brustein:2007jj}
R.~Brustein, D.~Gorbonos and M.~Hadad,
%``Wald's entropy is equal to a quarter of the horizon area in units of the effective gravitational coupling,''
Phys. Rev. D \textbf{79}, 044025 (2009)
doi:10.1103/PhysRevD.79.044025
[arXiv:0712.3206 [hep-th]].
%104 citations counted in INSPIRE as of 09 Jan 2021

		

%\cite{Banados:1993ur}
\bibitem{Banados:1993ur}
M.~Banados, C.~Teitelboim and J.~Zanelli,
%``Dimensionally continued black holes,''
Phys. Rev. D \textbf{49}, 975-986 (1994)
doi:10.1103/PhysRevD.49.975
[arXiv:gr-qc/9307033 [gr-qc]].
%221 citations counted in INSPIRE as of 02 Dec 2020		
		
		
%\cite{Fan:2016zfs}
\bibitem{Fan:2016zfs}
Z.~Y.~Fan, B.~Chen and H.~Lu,
%``Criticality in Einstein\textendash{}Gauss\textendash{}Bonnet gravity: gravity without graviton,''
Eur. Phys. J. C \textbf{76}, no.10, 542 (2016)
doi:10.1140/epjc/s10052-016-4389-x
[arXiv:1606.02728 [hep-th]].
%20 citations counted in INSPIRE as of 29 Nov 2020	
	






%\cite{Stanford:2014jda}
\bibitem{Stanford:2014jda} 
D.~Stanford and L.~Susskind,
%``Complexity and Shock Wave Geometries,''
Phys.\ Rev.\ D {\bf 90}, no. 12, 126007 (2014)
doi:10.1103/PhysRevD.90.126007
[arXiv:1406.2678 [hep-th]].
%%CITATION = doi:10.1103/PhysRevD.90.126007;%%
%111 citations counted in INSPIRE as of 08 Mar 2018


%\cite{Brown:2015bva}
\bibitem{Brown:2015bva} 
A.~R.~Brown, D.~A.~Roberts, L.~Susskind, B.~Swingle and Y.~Zhao,
%``Holographic Complexity Equals Bulk Action?,''
Phys.\ Rev.\ Lett.\  {\bf 116}, no. 19, 191301 (2016)
doi:10.1103/PhysRevLett.116.191301
[arXiv:1509.07876 [hep-th]].
%%CITATION = doi:10.1103/PhysRevLett.116.191301;%%
%137 citations counted in INSPIRE as of 03 Apr 2018	
	
	
	%\cite{Brown:2015lvg}
\bibitem{Brown:2015lvg} 
A.~R.~Brown, D.~A.~Roberts, L.~Susskind, B.~Swingle and Y.~Zhao,
%``Complexity, action, and black holes,''
Phys.\ Rev.\ D {\bf 93}, no. 8, 086006 (2016)
doi:10.1103/PhysRevD.93.086006
[arXiv:1512.04993 [hep-th]].
%%CITATION = doi:10.1103/PhysRevD.93.086006;%%
%104 citations counted in INSPIRE as of 14 Mar 2018

%\cite{Carmi:2016wjl}
\bibitem{Carmi:2016wjl} 
D.~Carmi, R.~C.~Myers and P.~Rath,
%``Comments on Holographic Complexity,''
JHEP {\bf 1703}, 118 (2017)
doi:10.1007/JHEP03(2017)118
[arXiv:1612.00433 [hep-th]].
%%CITATION = doi:10.1007/JHEP03(2017)118;%%
%61 citations counted in INSPIRE as of 28 Apr 2018

%\cite{Yang:2016awy}
\bibitem{Yang:2016awy}
R.~Q.~Yang,
%``Strong energy condition and complexity growth bound in holography,''
Phys. Rev. D \textbf{95}, no.8, 086017 (2017)
doi:10.1103/PhysRevD.95.086017
[arXiv:1610.05090 [gr-qc]].
%55 citations counted in INSPIRE as of 29 Nov 2020







%\cite{Carmi:2017jqz}
\bibitem{Carmi:2017jqz} 
D.~Carmi, S.~Chapman, H.~Marrochio, R.~C.~Myers and S.~Sugishita,
%``On the Time Dependence of Holographic Complexity,''
JHEP {\bf 1711}, 188 (2017)
doi:10.1007/JHEP11(2017)188
[arXiv:1709.10184 [hep-th]].
%%CITATION = doi:10.1007/JHEP11(2017)188;%%
%22 citations counted in INSPIRE as of 08 Mar 2018	





%\cite{Kim:2017qrq}
\bibitem{Kim:2017qrq}
R.~Q.~Yang, C.~Niu, C.~Y.~Zhang and K.~Y.~Kim,
%``Comparison of holographic and field theoretic complexities for time dependent thermofield double states,''
JHEP \textbf{02}, 082 (2018)
doi:10.1007/JHEP02(2018)082
[arXiv:1710.00600 [hep-th]].
%89 citations counted in INSPIRE as of 29 Nov 2020




%\cite{Yang:2017nfn}
\bibitem{Yang:2017nfn}
R.~Q.~Yang,
%``Complexity for quantum field theory states and applications to thermofield double states,''
Phys. Rev. D \textbf{97}, no.6, 066004 (2018)
doi:10.1103/PhysRevD.97.066004
[arXiv:1709.00921 [hep-th]].
%64 citations counted in INSPIRE as of 29 Nov 2020


%\cite{Kim:2017lrw}
\bibitem{Kim:2017lrw} 
R.~Q.~Yang, C.~Niu and K.~Y.~Kim,
%``Surface Counterterms and Regularized Holographic Complexity,''
JHEP {\bf 1709}, 042 (2017)
doi:10.1007/JHEP09(2017)042
[arXiv:1701.03706 [hep-th]].
%%CITATION = doi:10.1007/JHEP09(2017)042;%%
%17 citations counted in INSPIRE as of 28 Apr 2018





%\cite{Miao:2017quj}
\bibitem{Miao:2017quj} 
Y.~G.~Miao and L.~Zhao,
%``Complexity-action duality of the shock wave geometry in a massive gravity theory,''
Phys.\ Rev.\ D {\bf 97}, no. 2, 024035 (2018)
doi:10.1103/PhysRevD.97.024035
[arXiv:1708.01779 [hep-th]].
%%CITATION = doi:10.1103/PhysRevD.97.024035;%%
%4 citations counted in INSPIRE as of 28 Apr 2018	
	







%\cite{Yang:2018nda}
\bibitem{Yang:2018nda}
R.~Q.~Yang, Y.~S.~An, C.~Niu, C.~Y.~Zhang and K.~Y.~Kim,
%``Principles and symmetries of complexity in quantum field theory,''
Eur. Phys. J. C \textbf{79}, no.2, 109 (2019)
doi:10.1140/epjc/s10052-019-6600-3
[arXiv:1803.01797 [hep-th]].
%53 citations counted in INSPIRE as of 29 Nov 2020





%\cite{Chen:2018mcc}
\bibitem{Chen:2018mcc}
B.~Chen, W.~M.~Li, R.~Q.~Yang, C.~Y.~Zhang and S.~J.~Zhang,
%``Holographic subregion complexity under a thermal quench,''
JHEP \textbf{07}, 034 (2018)
doi:10.1007/JHEP07(2018)034
[arXiv:1803.06680 [hep-th]].
%57 citations counted in INSPIRE as of 29 Nov 2020

%\cite{An:2018dbz}
\bibitem{An:2018dbz}
Y.~S.~An, R.~G.~Cai and Y.~Peng,
%``Time Dependence of Holographic Complexity in Gauss-Bonnet Gravity,''
Phys. Rev. D \textbf{98}, no.10, 106013 (2018)
doi:10.1103/PhysRevD.98.106013
[arXiv:1805.07775 [hep-th]].
%34 citations counted in INSPIRE as of 29 Nov 2020


%\cite{Guo:2018kzl}
\bibitem{Guo:2018kzl}
M.~Guo, J.~Hernandez, R.~C.~Myers and S.~M.~Ruan,
%``Circuit Complexity for Coherent States,''
JHEP \textbf{10}, 011 (2018)
doi:10.1007/JHEP10(2018)011
[arXiv:1807.07677 [hep-th]].
%70 citations counted in INSPIRE as of 05 Dec 2020


%\cite{Yang:2018tpo}
\bibitem{Yang:2018tpo}
R.~Q.~Yang, Y.~S.~An, C.~Niu, C.~Y.~Zhang and K.~Y.~Kim,
%``More on complexity of operators in quantum field theory,''
JHEP \textbf{03}, 161 (2019)
doi:10.1007/JHEP03(2019)161
[arXiv:1809.06678 [hep-th]].
%35 citations counted in INSPIRE as of 29 Nov 2020






%\cite{Yang:2018cgx}
\bibitem{Yang:2018cgx}
R.~Q.~Yang and K.~Y.~Kim,
%``Complexity of operators generated by quantum mechanical Hamiltonians,''
JHEP \textbf{03}, 010 (2019)
doi:10.1007/JHEP03(2019)010
[arXiv:1810.09405 [hep-th]].
%11 citations counted in INSPIRE as of 29 Nov 2020




%\cite{Yang:2019gce}
\bibitem{Yang:2019gce}
R.~Q.~Yang, H.~S.~Jeong, C.~Niu and K.~Y.~Kim,
%``Complexity of Holographic Superconductors,''
JHEP \textbf{04}, 146 (2019)
doi:10.1007/JHEP04(2019)146
[arXiv:1902.07586 [hep-th]].
%21 citations counted in INSPIRE as of 29 Nov 2020

%\cite{Bernamonti:2019zyy}
\bibitem{Bernamonti:2019zyy}
A.~Bernamonti, F.~Galli, J.~Hernandez, R.~C.~Myers, S.~M.~Ruan and J.~Sim\'on,
%``First Law of Holographic Complexity,''
Phys. Rev. Lett. \textbf{123}, no.8, 081601 (2019)
doi:10.1103/PhysRevLett.123.081601
[arXiv:1903.04511 [hep-th]].
%38 citations counted in INSPIRE as of 05 Dec 2020


%\cite{Yang:2019iav}
\bibitem{Yang:2019iav}
R.~Q.~Yang and K.~Y.~Kim,
%``Time evolution of the complexity in chaotic systems: a concrete example,''
JHEP \textbf{05}, 045 (2020)
doi:10.1007/JHEP05(2020)045
[arXiv:1906.02052 [hep-th]].
%13 citations counted in INSPIRE as of 29 Nov 2020

	
%\cite{Yang:2019udi}
\bibitem{Yang:2019udi}
R.~Q.~Yang, Y.~S.~An, C.~Niu, C.~Y.~Zhang and K.~Y.~Kim,
%``To be unitary-invariant or not?: a simple but non-trivial proposal for the complexity between states in quantum mechanics/field theory,''
[arXiv:1906.02063 [hep-th]].
%13 citations counted in INSPIRE as of 29 Nov 2020	

%\cite{Caceres:2019pgf}
\bibitem{Caceres:2019pgf}
E.~Caceres, S.~Chapman, J.~D.~Couch, J.~P.~Hernandez, R.~C.~Myers and S.~M.~Ruan,
%``Complexity of Mixed States in QFT and Holography,''
JHEP \textbf{03}, 012 (2020)
doi:10.1007/JHEP03(2020)012
[arXiv:1909.10557 [hep-th]].
%27 citations counted in INSPIRE as of 05 Dec 2020
	
%\cite{An:2019opz}
\bibitem{An:2019opz}
Y.~S.~An, R.~G.~Cai, L.~Li and Y.~Peng,
%``Holographic complexity growth in an FLRW universe,''
Phys. Rev. D \textbf{101}, no.4, 046006 (2020)
doi:10.1103/PhysRevD.101.046006
[arXiv:1909.12172 [hep-th]].
%9 citations counted in INSPIRE as of 29 Nov 2020	
	
%\cite{Yang:2019alh}
\bibitem{Yang:2019alh}
R.~Q.~Yang,
%``Upper bound on cross sections inside black holes and complexity growth rate,''
Phys. Rev. D \textbf{102}, no.10, 106001 (2020)
doi:10.1103/PhysRevD.102.106001
[arXiv:1911.12561 [hep-th]].
%3 citations counted in INSPIRE as of 29 Nov 2020

%\cite{Bernamonti:2020bcf}
\bibitem{Bernamonti:2020bcf}
A.~Bernamonti, F.~Galli, J.~Hernandez, R.~C.~Myers, S.~M.~Ruan and J.~Sim\'on,
%``Aspects of The First Law of Complexity,''
doi:10.1088/1751-8121/ab8e66
[arXiv:2002.05779 [hep-th]].
%16 citations counted in INSPIRE as of 05 Dec 2020


%\cite{Ruan:2020vze}
\bibitem{Ruan:2020vze}
S.~M.~Ruan,
%``Purification Complexity without Purifications,''
[arXiv:2006.01088 [hep-th]].
%1 citations counted in INSPIRE as of 05 Dec 2020


%\cite{An:2020tkn}
\bibitem{An:2020tkn}
Y.~S.~An, T.~Ji and L.~Li,
%``Magnetotransport and Complexity of Holographic Metal-Insulator Transitions,''
JHEP \textbf{10}, 023 (2020)
doi:10.1007/JHEP10(2020)023
[arXiv:2007.13918 [hep-th]].
%2 citations counted in INSPIRE as of 29 Nov 2020	
	
%\cite{Hernandez:2020nem}
\bibitem{Hernandez:2020nem}
J.~Hernandez, R.~C.~Myers and S.~M.~Ruan,
%``Quantum Extremal Islands Made Easy, PartIII: Complexity on the Brane,''
[arXiv:2010.16398 [hep-th]].
%3 citations counted in INSPIRE as of 05 Dec 2020

%\cite{Yang:2020tna}
\bibitem{Yang:2020tna}
R.~Q.~Yang, Y.~S.~An, C.~Niu, C.~Y.~Zhang and K.~Y.~Kim,
%``What kind of ''complexity'' is dual to holographic complexity?,''
[arXiv:2011.14636 [hep-th]].
%0 citations counted in INSPIRE as of 02 Dec 2020
		


		
	\end{thebibliography}
\end{document}